\documentclass[aps,prd,preprintnumbers,superscriptaddress,nofootinbib,notitlepage,floatfix,10pt]{revtex4-2}
\usepackage[pdftex]{graphicx}
\usepackage{bm,latexsym,amsmath,amssymb,amsfonts,mathrsfs}
\usepackage{color}
\allowdisplaybreaks[1]
\usepackage[pdftex,colorlinks=true,linkcolor=blue,citecolor=cyan,backref=page]{hyperref}
\newcommand*{\D}{\mathrm{d}}
\newcommand*{\mpl}{M_{\mathrm{Pl}}}
\newcommand{\wB}{\widetilde{B}}
\begin{document}
%-----------------------------------------------------------------%
\title{Ghost-free 2-form fields in cosmology: implications for gravitational parity violation}
%-----------------------------------------------------------------%
%
\author{Yuki~Horii}
\email[Email: ]{yuki.horii@rikkyo.ac.jp}
\affiliation{Department of Physics, Rikkyo University, Toshima, Tokyo 171-8501, Japan}
\author{Tomoaki~Murata}
\email[Email: ]{tomoaki-m@metro-cit.ac.jp}
\affiliation{Department of General Education, Tokyo Metropolitan College of Industrial Technology, Shinagawa, Tokyo 140-0011, Japan}
\affiliation{Department of Physics, Rikkyo University, Toshima, Tokyo 171-8501, Japan}
\author{Tsutomu~Kobayashi}
\email[Email: ]{tsutomu@rikkyo.ac.jp}
\affiliation{Department of Physics, Rikkyo University, Toshima, Tokyo 171-8501, Japan}
%
%-----------------------------------------------------------%
\begin{abstract}
We explore the possibility of parity-violating, nonminimally coupled 2-form field theories that retain the same dynamical degrees of freedom as a massive 2-form and thus are ghost-free.
Starting from the most general kinetic terms and dimension four couplings between the 2-form field and the curvature tensors, we find a two-parameter family of such theories.
However, we also find that parity-violating terms involving the dual 2-form field can be absorbed into a field redefinition, leaving a theory with essentially the same structure as that obtained by Heisenberg and Trenkler (i.e., the parity-preserving coupling to the double dual Riemann tensor).
After the field redefinition, the only place where parity-violating terms can appear is in the potential.
We then consider a homogeneous and isotropic cosmological model with an isotropic configuration of a triplet of 2-form fields, and study tensor perturbations in this setup.
There are three types of tensor perturbations, two of which are dynamical.
We show that chiral gravitational waves can be generated in the presence of parity-violating terms in the potential.
\end{abstract}
%-----------------------------------------------------------------%
\preprint{RUP-25-25}
\maketitle
%-----------------------------------------------------------------%

\section{Introduction}\label{sec:intro}

The direct detection of gravitational waves by the LIGO--Virgo--KAGRA collaboration has opened a new window into gravity in the strong-field regime, offering us new observational tests of general relativity~\cite{LIGOScientific:2018mvr,LIGOScientific:2020ibl,KAGRA:2021vkt,LIGOScientific:2025slb}.
Measurements of the polarization and propagation properties of gravitational waves allow us to access information on the underlying dynamical degrees of freedom and symmetries of gravity.
Among various possibilities, parity violation in the gravity sector is one of the most intriguing signals beyond general relativity.
It arises in a variety of fundamental or effective field theories, and a sizable amount of chiral gravitational waves can be produced in different cosmological scenarios.

Parity violation has often been studied in the context of modified gravity theories.
A prototypical example is Chern--Simons gravity, in which a scalar (axion-like) field couples to the Pontryagin density $R\tilde{R}$ and induces parity-odd modifications to the propagation of gravitational waves~\cite{Jackiw:2003pm,Alexander:2009tp} (see Ref.~\cite{Sulantay:2022sag} for a discussion on parity breaking in the Palatini formulation of Chern--Simons gravity).
More recently, higher-derivative operators that violate parity while remaining ghost-free have been systematically classified within generalized scalar-tensor theories~\cite{Crisostomi:2017ugk}.
{Parity violation can also occur in Ho\v{r}ava--Lifshitz gravity~\cite{Takahashi:2009wc,Zhu:2013fja} and in gravity based on non-Riemannian geometry~\cite{Conroy:2019ibo,Li:2020xjt,Li:2021wij,Li:2022mti,Wu:2021ndf,Hohmann:2022wrk}.
Meanwhile, model-independent analyses have clarified the generic structure of parity-odd contributions to the gravitational-wave dispersion relation, as well as their implications for observations~\cite{Nishizawa:2018srh,Jenks:2023pmk}.

In the context of (axion-)gauge field inflation, chiral gravitational waves can naturally be generated due to mixing with parity-breaking extra tensor modes~\cite{Adshead:2013nka,Maleknejad:2016qjz,Dimastrogiovanni:2016fuu,Watanabe:2020ctz,Iarygina:2021bxq,Dimastrogiovanni:2023oid,Murata:2024urv} (see also Refs.~\cite{Bielefeld:2015daa,Tishue:2021blv}).
A model-independent effective field theory encompassing these models has been developed in Ref.~\cite{Aoki:2025uwz}.
An effective field theory study of parity violation in gravitational waves from an extra spectator tensor field has been carried out in Ref.~\cite{Garriga:2025uko}.
Furthermore, it has been pointed out recently that antisymmetric rank-2 tensor fields (2-form fields) can induce parity violation in gravitational waves through nonminimal coupling to gravity~\cite{Manton:2024hyc,Alexander:2025wnj}.

In this paper, we focus on the antisymmetric 2-form field $B_{\mu\nu}$ and discuss its relevance to gravitational parity violation.
A 2-form field transforms in the $(1,0)\oplus(0,1)$ representation of the four-dimensional Lorentz group and appears naturally in many effective field theories, most famously as the Kalb--Ramond field in string theory~\cite{PhysRevD.9.2273}, but also in a variety of other contexts~\cite{Terschlusen:2013iqa,Copeland:1984qk,Beekman:2010zx}.
The duality properties of a 2-form field depend sensitively on its mass and interactions: a massless free 2-form field is dual to a pseudoscalar, while a massive noninteracting 2-form field is dual to a Proca field.
However, a general interacting massive 2-form field possesses no such simple dual description~\cite{Capanelli:2023uwv}.

It is interesting to note that a 2-form field admits a parity-odd nonminimal coupling to the Riemann tensor already at mass dimension four~\cite{Manton:2024hyc,Alexander:2025wnj},
\begin{align}
    B^{\mu\nu}\wB^{\alpha\beta}R_{\mu\nu\alpha\beta},
\end{align}
allowing one to construct a low-energy effective theory that exhibits gravitational parity violation without invoking higher-dimensional operators.
Here, $\wB^{\mu\nu}$ is the dual 2-form field defined by
\begin{align}
    \wB^{\alpha\beta}:=\frac{1}{2\sqrt{-g}}\epsilon^{\alpha\beta\mu\nu}B_{\mu\nu},
\end{align}
where $\epsilon^{\alpha\beta\mu\nu}$ is the totally antisymmetric symbol with $\epsilon^{0123}=+1$ and $g$ is the determinant of the metric, $g=\textrm{det}(g_{\mu\nu})$.
This coupling is a part of the general action for a nonminimally coupled 2-form field 
displayed by Altschul \textit{et al.}~\cite{Altschul:2009ae}.
They systematically constructed the most general kinetic terms for the 2-form field and dimension four couplings between the 2-form field and the curvature tensors, demonstrating that both parity-even and parity-odd structures are included.
However, the general theory given in Ref.~\cite{Altschul:2009ae} contains more propagating degrees of freedom than the massive 2-form field theory, which would be problematic from the viewpoint of ghost instabilities.
Meanwhile, by requiring that the equations of motion remain of second order
and hence the resultant theory retains the same dynamical degrees of freedom as a massive 2-form,
Heisenberg and Trenkler showed that the allowed \textit{parity-even} coupling is 
of the form $\sim B\cdot B\cdot L$, where $L$ represents the double dual Riemann tensor~\cite{Heisenberg:2019akx}.
In this paper, we start from the general action of Altschul \textit{et al.}~\cite{Altschul:2009ae}
including \textit{parity-odd} terms (i.e., terms involving $\wB^{\mu\nu}$), and single out the allowed terms that yield no extra propagating degrees of freedom.

After driving a healthy 2-form field theory, we study cosmology in the presence of 2-form fields.
A single 2-form field sources anisotropies in general.
To implement 2-form fields in homogeneous and isotropic cosmology, 
we introduce a triplet of 2-form fields, in analogy with a triplet of mutually orthogonal vector fields.
In this setup, we consider tensor perturbations, which arise from the triplet of the 2-form fields as well as the metric, and explore under which circumstances gravitational parity violation occurs and chiral gravitational waves can be generated.

The structure of the paper is as follows.
In Sec.~II, 
we determine the healthy subset of nonminimally coupled 2-form field theories starting from
the general action of Altschul \textit{et al.}~\cite{Altschul:2009ae}.
We then introduce a triplet of the 2-form fields and derive the basic equations governing homogeneous and isotropic cosmology in Sec.~III.
In Sec.~IV, we analyze tensor perturbations in this triplet setup and identify under which circumstances gravitational parity violation occurs.
Finally, we draw our conclusions in Sec.~V.

\if0
Appendix~A summarizes the analogous ghost-free conditions for vector fields and the associated constraints on nonminimal couplings.
Appendix~B provides explicit expressions for the coefficients in the tensor perturbation action presented in Sec.~IV.
\fi

\section{Healthy 2-form field theory}

In this section, we consider the general kinetic terms for a 2-form field
and dimension four couplings between the 2-form field and the curvature tensors~\cite{Altschul:2009ae} (see also~\cite{Seifert:2019kuz,Potting:2023xzt}).
To determine the instability-free subset, we first switch off gravity and
study the stability of the 2-form field with the general kinetic terms,
and then include the dynamics of gravity to single out a healthy nonminimal coupling.
The latter analysis is done in a spatially flat and homogeneous cosmological setup
for simplicity.
As shown in Appendix~\ref{app:vector}, a healthy theory of a nonminimally coupled vector field can also be derived in this way, which is nothing but a subset of the generalized Proca theory.

\subsection{The starting action}

We start with exhausting all possible kinetic terms and dimension-four operators coupling the 2-form field to curvature tensors.
We then show that some of them can be removed without loss of generality by the use of some identities and integration by parts, following closely Ref.~\cite{Altschul:2009ae}.

Let us first consider the kinetic terms.
An important piece for the construction of the kinetic terms is the field strength tensor,
\begin{align}
    H_{\mu\nu\rho}= \partial_{\mu}B_{\nu\rho}+\partial_{\nu}B_{\rho\mu}+\partial_{\rho}B_{\mu\nu}
    =3\partial_{[\mu}B_{\nu\rho]}.
\end{align}
The field strength tensor is invariant under the gauge transformation
$B_{\mu\nu}\to B_{\mu\nu}+\partial_\mu V_\nu-\partial_\nu V_\mu$,
where $V_\mu$ is an arbitrary vector field.
The standard gauge-invariant kinetic term is given by
\begin{align}
    -\frac{1}{12}H_{\mu\nu\rho}H^{\mu\nu\rho}
    =-\frac{1}{4}\nabla_{\rho}B_{\mu\nu}\nabla^{\rho}B^{\mu\nu}
    -\frac{1}{2}\nabla_{\mu}B_{\nu\rho}\nabla^{\rho}B^{\mu\nu}.
    \label{eq:H^2-explicit}
\end{align}
However, since we no longer assume the gauge symmetry, we may consider all of the following terms as possible kinetic terms:
\begin{align}
    &\nabla_{\rho}B_{\mu\nu}\nabla^{\rho}B^{\mu\nu}, \quad
    \nabla_{\mu}B_{\nu\rho}\nabla^{\rho}B^{\mu\nu}, \quad
    \nabla_{\mu}B^{\mu\rho}\nabla^{\nu}B_{\nu\rho},
    \notag\\
    &\nabla_{\rho}B_{\mu\nu}\nabla^{\rho}\wB^{\mu\nu}, \quad
    \nabla_{\mu}B_{\nu\rho}\nabla^{\rho}\wB^{\mu\nu}, \quad
    \nabla_{\mu}B^{\mu\rho}\nabla^{\nu}\wB_{\nu\rho}.
    \label{possible terms}
\end{align}
Noting the relations $\nabla_{\mu}B_{\nu\rho}\nabla^{\nu}B^{\mu\rho}=-\nabla_{\mu}B_{\nu\rho}\nabla^{\rho}B^{\mu\nu}$ and $\nabla_{\mu}B_{\nu\rho}\nabla^{\nu}\wB^{\mu\rho}=-\nabla_{\mu}B_{\nu\rho}\nabla^{\rho}\wB^{\mu\nu}$, we have already removed $\nabla_{\mu}B_{\nu\rho}\nabla^{\nu}B^{\mu\rho}$ and $\nabla_{\mu}B_{\nu\rho}\nabla^{\nu}\wB^{\mu\rho}$ from the above list.
As seen from Eq.~\eqref{eq:H^2-explicit}, one may take the standard kinetic term as an independent term in place of $\nabla_{\rho}B_{\mu\nu}\nabla^{\rho}B^{\mu\nu}$.

Let us comment on the point that was overlooked in Ref.~\cite{Altschul:2009ae}.
It can be shown that the following identity holds:
\begin{equation}
    \nabla_{(\alpha}B_{\mu\lambda}\nabla_{\beta)}\wB^{\lambda\nu}=-\frac{1}{4}\delta^{\nu}_{\mu}\nabla_{\alpha}B_{\rho\lambda}\nabla_{\beta}\wB^{\rho\lambda}.
    \label{identity 1}
\end{equation}
Contracting this with $g^{\alpha\mu}\delta_\beta^\nu$, we obtain
\begin{equation}
    \nabla_{\mu}B_{\nu\rho}\nabla^{\mu}\wB^{\nu\rho}
    +2\nabla_{\mu}B_{\nu\rho}\nabla^{\rho}\wB^{\mu\nu}
    -2\nabla_{\mu}B^{\mu\rho}\nabla^{\nu}\wB_{\nu\rho}
    =0.
    \label{identity 2}
\end{equation}
Using this relation, one can remove the sixth term in Eq.~\eqref{possible terms}.

Summarizing the above, the independent kinetic terms are
\begin{equation}
    -\frac{1}{12}H_{\mu\nu\rho}H^{\mu\nu\rho}, \quad \nabla_{\mu}B_{\nu\rho}\nabla^{\rho}B^{\mu\nu}, \quad \nabla_{\mu}B^{\mu\rho}\nabla^{\nu}B_{\nu\rho}, \quad \nabla_{\mu}B_{\nu\rho}\nabla^{\mu}\wB^{\nu\rho}, \quad \nabla_{\mu}B_{\nu\rho}\nabla^{\rho}\wB^{\mu\nu}.
    \label{eq:kinetic-term-list}
\end{equation}

Let us next consider dimension-four operators coupling $B_{\mu\nu}$ with the curvature tensors.
The possible couplings are
\begin{equation}
    X_BR, \quad Y_BR, \quad B^{\mu\lambda}B_{\lambda}^{\ \nu}R_{\mu\nu}, \quad B^{\mu\lambda}\wB_{\lambda}^{\ \nu}R_{\mu\nu}, \quad B^{\mu\nu}B^{\alpha\beta}R_{\mu\nu\alpha\beta}, \quad B^{\mu\nu}\wB^{\alpha\beta}R_{\mu\nu\alpha\beta},
    \label{non-minimal coupling terms 1}
\end{equation}
where $X_B$ and $Y_B$ are defined as
\begin{equation}
    X_B= B_{\mu\nu}B^{\mu\nu}, \quad Y_B= B_{\mu\nu}\wB^{\mu\nu}.
\end{equation}
Concerning the coupling to the Riemann tensor, one may notice that indices can be contracted in several different ways.
However, we have the relations
\begin{align}
    (B^{\mu\nu}B^{\alpha\beta}-2B^{\mu\alpha}B^{\nu\beta})R_{\mu\nu\alpha\beta}
    &
    =3 B^{\mu\nu}B^{\alpha\beta}R_{[\mu\nu\alpha]\beta}=0,
    \notag \\
    (B^{\mu\nu}\wB^{\alpha\beta}-2B^{\mu\alpha}\wB^{\nu\beta})
    R_{\mu\nu\alpha\beta}
    &=3 B^{\mu\nu}\wB^{\alpha\beta}R_{[\mu\nu\alpha]\beta}=0,
    \notag \\
    (B^{\mu\nu}B^{\alpha\beta}-2B^{\mu\alpha}B^{\nu\beta})
    \widetilde{R}_{\mu\nu\alpha\beta}
    &=3 B^{\mu\nu}B^{\alpha\beta}\widetilde{R}_{[\mu\nu\alpha]\beta}
    =\frac{5}{6} B^{\mu\nu}B^{\alpha\beta}\varepsilon_{\mu\nu\alpha\lambda}R^{\lambda}_{\ \beta}
    =\frac{5}{3}B^{\alpha\beta}\wB_{\lambda\alpha}R^{\lambda}_{\ \beta},
\end{align}
and therefore it is sufficient to consider the two types of coupling to the Riemann tensor presented in Eq.~\eqref{non-minimal coupling terms 1}.

One can further remove three terms from the list of Eq.~\eqref{non-minimal coupling terms 1}.
First, there is an identity,
\begin{equation}
    B_{\mu\lambda}\wB^{\lambda\nu}=-\frac{1}{4}\delta^{\nu}_{\mu}Y_B,
\end{equation}
and therefore we do not need to consider the fourth term in Eq.~\eqref{non-minimal coupling terms 1}.
Furthermore, one can show that the following combinations are total derivatives:
\begin{align}
    &\nabla_{\mu}B_{\nu\rho}\nabla^{\rho}B^{\mu\nu}
    +\nabla_{\mu}B^{\mu\rho}\nabla^{\nu}B_{\nu\rho}
    +B_{\mu\lambda}B^{\lambda\nu}R^{\mu}_{\ \nu}
    +\frac{1}{2}B^{\mu\nu}B^{\alpha\beta}R_{\mu\nu\alpha\beta},
    \\
    &\nabla_{\mu}B_{\nu\rho}\nabla^{\rho}\wB^{\mu\nu}
    +\nabla_{\mu}B^{\mu\rho}\nabla^{\nu}\wB_{\nu\rho}
    -\frac{1}{4}Y_BR
    +\frac{1}{2}B^{\mu\nu}\wB^{\alpha\beta}R_{\mu\nu\alpha\beta}.
\end{align}
Therefore, we do not need to keep the last two terms in Eq.~\eqref{non-minimal coupling terms 1}, as long as we consider all the kinetic terms in Eq.~\eqref{eq:kinetic-term-list}.
In particular, the nonminimal coupling considered in Ref.~\cite{Manton:2024hyc} can be recast into the kinetic terms and the nonminimal coupling $Y_BR$.

In summary, our starting action is
\begin{align}
    S
    &=\int \D^4x\sqrt{-g}\bigg[
    \frac{M_{\text{Pl}}^2}{2}R
    -\frac{1}{12}H_{\mu\nu\rho}H^{\mu\nu\rho}
    +\frac{\eta_1}{2}\nabla_{\mu}B_{\nu\rho}\nabla^{\rho}B^{\mu\nu}
    +\frac{\eta_2}{2}\nabla_{\mu}B^{\mu\rho}\nabla^{\nu}B_{\nu\rho}
    +\frac{\eta_3}{2}\nabla_{\mu}B_{\nu\rho}\nabla^{\mu}\wB^{\nu\rho}
    \notag \\& \quad 
    +\frac{\eta_4}{2}\nabla_{\mu}B_{\nu\rho}\nabla^{\rho}\wB^{\mu\nu}
    -V(X_B,Y_B)
    +\frac{\xi_1}{2}X_BR
    +\frac{\xi_2}{2}B^{\mu\lambda}B_{\lambda\nu}R^{\nu}_{\ \mu}
    +\frac{\xi_3}{2}Y_BR
    \bigg],
    \label{general action 1}
\end{align}
where $\eta_i$ and $\xi_i$ are constants, and we have included the Einstein-Hilbert term and the potential $V(X,Y)$.
The coefficient of the standard kinetic term could be arbitrary, but we adopt the canonical normalization for simplicity, assuming that it is nonvanishing.
The difference from the action presented in~\cite{Altschul:2009ae} is that a redundant kinetic term is removed by the use of the identity~\eqref {identity 2}.
The other differences are only apparent and amount to total derivatives.

\subsection{Analysis in the Minkowski background}

The theory described by the action~\eqref{general action 1} contains unstable degrees of freedom in general.
To derive the conditions that remove the dangerous degrees of freedom, we start by switching off gravity and consider a 2-form field in Minkowski,
\begin{align}
    S
    &=\int \D^4x\left[
    -\frac{1}{12}H_{\mu\nu\rho}H^{\mu\nu\rho}
    +\frac{\eta_1}{2}\partial_{\mu}B_{\nu\rho}\partial^{\rho}B^{\mu\nu}
    +\frac{\eta_2}{2}\partial_{\mu}B^{\mu\rho}\partial^{\nu}B_{\nu\rho}
    +\frac{\eta_3}{2}\partial_{\mu}B_{\nu\rho}\partial^{\mu}\wB^{\nu\rho}
    +\frac{\eta_4}{2}\partial_{\mu}B_{\nu\rho}\partial^{\rho}\wB^{\mu\nu}
    \right]
    \notag \\ &=
    \int \D^4x\left[
    -\frac{1}{12}H_{\mu\nu\rho}H^{\mu\nu\rho}
    +\frac{\eta_1-\eta_2}{2}\partial_{\mu}B_{\nu\rho}\partial^{\rho}B^{\mu\nu}
    %+\frac{\eta_2}{2}\partial_{\mu}B^{\mu\rho}\partial^{\nu}B_{\nu\rho}
    +\frac{\eta_3}{2}\partial_{\mu}B_{\nu\rho}\partial^{\mu}\wB^{\nu\rho}
    +\frac{\eta_4}{2}\partial_{\mu}B_{\nu\rho}\partial^{\rho}\wB^{\mu\nu}
    \right].
    \label{action in Minkowski 1}
\end{align}
We ignore the potential for the moment and focus on the structure of the kinetic terms.

We introduce three-vectors $\mathcal{E}_{i}(t,\vec{x})$ and $\mathcal{M}_{i}(t,\vec{x})$ defined as
\begin{equation}
    \mathcal{E}_{i}=B_{i0}, \quad 
    \mathcal{M}_{i}=\frac{1}{2}\epsilon_{ijk}
    B_{jk},
    \label{decompose 0i and ij}
\end{equation}
in terms of which the action~\eqref{action in Minkowski 1} is written as
\begin{align}
    S
    &=\int \D^4x\biggl\{
    \frac{1}{2}\left[
    \dot{\mathcal{M}}_{i}\dot{\mathcal{M}}_{i}
    -2\alpha\dot{\mathcal{M}}_{i}\dot{\mathcal{E}}_{i}
    +(\alpha^2+\gamma)\dot{\mathcal{E}}_{i}\dot{\mathcal{E}}_{i}
    \right]
    +(1+\alpha^2+\gamma)\epsilon_{ijk}\dot{\mathcal{M}}_{i}\partial_{j}\mathcal{E}_{k}
    \notag\\
    &\quad 
    -\frac{1}{2}(1+\alpha^2+\gamma)\partial_i\mathcal{E}_{i}\partial_{j}\mathcal{E}_{j}
    +\frac{1}{2}\partial_{i}\mathcal{E}_{j}\partial_{i}\mathcal{E}_{j}
    +\alpha \partial_{i}\mathcal{E}_{j}\partial_{i}\mathcal{M}_{j}
    \notag\\
    &\quad 
    -\frac{1}{2}(1+\alpha^2+\gamma)\partial_{i}\mathcal{M}_{i}\partial_{j}\mathcal{M}_{j}
    +\frac{1}{2}(\alpha^2+\gamma)\partial_{i}\mathcal{M}_{j}\partial_{i}\mathcal{M}_{j}\biggr\},
    \label{action in Minkowski 2}
\end{align}
where $\epsilon_{ijk}$ is the totally antisymmetric symbol with $\epsilon_{123}=1$, a dot denotes differentiation with respect to $t$, and we defined
\begin{align}
    \alpha:=-\frac{1}{2}(4\eta_3-\eta_4),
    \quad 
    \gamma:=-\eta_1+\eta_2-\alpha^2.
\end{align}
We further split $\mathcal{E}_{i}$ and $\mathcal{M}_{i}$ into the longitudinal and transverse parts as
\begin{equation}
    \mathcal{E}_{i}=
    \partial_{i}\mathcal{E}^{\text{L}}+\mathcal{E}_{i}^{\text{T}},
    \quad
    \mathcal{M}_{i}=
    \partial_{i}\mathcal{M}^{\text{L}}+\mathcal{M}_{i}^{\text{T}},
    \label{deconpose longitudinal and transverse}
\end{equation}
where $\partial_{i}\mathcal{E}_{i}^{\text{T}}=0$ and $\partial_{i}\mathcal{M}_{i}^{\text{T}}=0$.
The action can now be written as $S=S^{\textrm{L}}+S^{\textrm{T}}$, with
\begin{align}
    S^{\textrm{L}}&=\frac{1}{2}\int\D^4 x\left[
        \dot\Phi^{\textrm{L}}(-\Delta)\dot\Phi^{\textrm{L}}
        -\Phi^{\textrm{L}}\Delta^2\Phi^{\textrm{L}}
        +\gamma\dot{\mathcal{E}}^{\textrm{L}}(-\Delta)\dot{\mathcal{E}}^{\textrm{L}}
        -\gamma\mathcal{E}^{\textrm{L}}\Delta^2\mathcal{E}^{\textrm{L}}
        \right],
\end{align}
and 
\begin{align}
    S^{\textrm{T}}=\frac{1}{2}\int\D^4x&\biggl[
        \left(\dot M_i^\textrm{T}+\epsilon_{ijk}\partial_jE_k^{\textrm{T}}\right)
        \left(\dot M_i^\textrm{T}+\epsilon_{ij'k'}\partial_{j'}E_{k'}^{\textrm{T}}\right)
        \notag \\ &
        +\frac{\gamma}{1+\alpha^2} \left(\dot E_i^{\textrm{T}}-\epsilon_{ijk}\partial_jM_k^{\textrm{T}}\right)
        \left(\dot E_i^{\textrm{T}}-\epsilon_{ij'k'}\partial_{j'}M_{k'}^{\textrm{T}}\right)
    \biggr],
\end{align}
where we performed the change of variables
\begin{align}
    \Phi^{\textrm{L}}:=\mathcal{M}^{\textrm{L}}
    -\alpha\mathcal{E}^{\textrm{L}},
    \quad  
    M_i^{\textrm{T}}:=\mathcal{M}_i^{\textrm{T}}
    -\alpha\mathcal{E}_i^{\textrm{T}},
    \quad 
    E_i^{\textrm{T}}:=\mathcal{E}_i^{\textrm{T}}
    +\alpha \mathcal{M}_i^{\textrm{T}}.
    \label{eq:redef-var-Mink}
\end{align}

It is clear that the longitudinal sector is healthy provided that $\gamma\ge 0$.
To see the pathology in the transverse sector, we rewrite the action by
introducing an auxiliary variable as 
\begin{align}
    S^{\textrm{T}}=\frac{1}{2}\int\D^4x&\biggl[-(\Delta\chi_i)^2+
        2\left(\dot M_i^\textrm{T}+\epsilon_{ijk}\partial_jE_k^{\textrm{T}}\right)
        \Delta\chi_i
        \notag \\ &
        +\frac{\gamma}{1+\alpha^2} \left(\dot E_i^{\textrm{T}}-\epsilon_{ijk}\partial_jM_k^{\textrm{T}}\right)
        \left(\dot E_i^{\textrm{T}}-\epsilon_{ij'k'}\partial_{j'}M_{k'}^{\textrm{T}}\right)
    \biggr].
\end{align}
Varying this action with respect to $M_i^{\textrm{T}}$, we obtain the constraint
\begin{align}
    \Delta \dot\chi_k+\frac{\gamma}{1+\alpha^2}\left(
        \epsilon_{i'j'k}\partial_{i'}\dot E_{j'}^{\textrm{T}}+\Delta M_k^{\textrm{T}}
    \right)=0.
\end{align}
This constraint can be used to remove $M_i^{\textrm{T}}$ from the action to get 
\begin{align}
    S^{\textrm{T}}=\frac{1}{2}\int\D^4x \left[
    -\frac{1+\alpha^2}{\gamma}\dot\chi_i(-\Delta)\dot\chi_i-(\Delta\chi_i)^2
    +2\dot\chi_i\dot V_i+2\Delta\chi_iV_i
    \right],
\end{align}
with $V_i:=\epsilon_{ijk}\partial_jE_k^{\textrm{T}}$.
Performing the change of variables $\chi_i=U_i-\gamma V_i/(1+\alpha^2)$, we arrive at
\begin{align}
    S^{\textrm{T}}=\frac{1}{2}\int\D^4x \left[
        \frac{\gamma}{1+\alpha^2}\dot V_i(-\Delta)\dot V_i
        -\frac{1+\alpha^2}{\gamma}\dot U_i(-\Delta)\dot U_i
        +(\Delta V_i)^2
        -\left(
            \Delta U_i-\frac{1+\alpha^2+\gamma}{1+\alpha^2}\Delta V_i
        \right)^2
    \right].
\end{align}
The signs of the kinetic terms are always opposite,
showing that one of the two fields $V_i$ and $U_i$ is a ghost.
This result forces us to consider the case of $\gamma=0$, i.e.,
\begin{align}
    -\eta_1+\eta_2-\frac{1}{4}(4\eta_3-\eta_4)^2=0.
    \label{eq:condition-detA=0}
\end{align}

In the case of $\gamma=0$, the action reads
\begin{align}
    S
    =\frac{1}{2}\int \D^4x \left[
    \dot\Phi^{\textrm{L}}(-\Delta)\dot\Phi^{\textrm{L}}
    -\Phi^{\textrm{L}}\Delta^2\Phi^{\textrm{L}}+
    \left(\dot M_i^\textrm{T}+\epsilon_{ijk}\partial_jE_k^{\textrm{T}}\right)
        \left(\dot M_i^\textrm{T}+\epsilon_{ij'k'}\partial_{j'}E_{k'}^{\textrm{T}}\right)
        -V
    \right],
\end{align}
where we restored the potential.
The $\eta_i$-dependence completely disappears from the action,
leaving a theory that retains the same dynamical degrees of freedom as a massive 2-form.
Note that $E_i^{\textrm{T}}$ is nondynamical.
We thus see that the condition~\eqref{eq:condition-detA=0} indeed removes the instability in the Minkowski background.

\subsection{Coupling to gravity}

The next step is to explore the consistent coupling of the 2-form field (and its dual) to gravity.
This can be done most efficiently by studying a homogeneous 2-form field in a spatially flat and homogeneous cosmological spacetime.
Although such an analysis does not retain full generality, we will find that the resultant conditions are sufficiently restrictive.

The metric and the 2-form field we consider are given, respectively, by
\begin{align}
    \D s^2=-N^2(t)\D t^2+\gamma_{ij}(t)\D x^{i}\D x^{j}
\end{align}
and 
\begin{align}
    B_{i0}=N\mathcal{E}_{i}(t), \quad B_{ij}=\sqrt{\gamma}\epsilon_{ijk}\mathcal{M}^{k}(t).
\end{align}
The field redefinition~\eqref{eq:redef-var-Mink} implies that the following combinations are useful:
\begin{align}
    M_i:=\mathcal{M}_i-\alpha \mathcal{E}_i,
    \quad 
    E_i:=\mathcal{E}_i+\alpha\mathcal{M}_i,
\end{align}
where the indices are raised and lowered with $\gamma_{ij}$.
Plugging the above expressions into the action~\eqref{general action 1} while imposing the condition~\eqref{eq:condition-detA=0}, we obtain
\begin{align}
    S
    &=\int \D^4x\sqrt{-g}\bigg[
    \frac{1}{2N^2}\mathcal{K}^{ij,kl}\dot{\gamma}_{ij}\dot{\gamma}_{kl}
    -V+\frac{1}{2N^2}\gamma^{ij}\dot M_i\dot M_j
    \notag \\ & \quad 
    +\frac{1}{2N^2}
    \left(\kappa_1\gamma^{ij}\gamma^{kl}
    +\kappa_2\gamma^{ik}\gamma^{jl}\right)
    E_i\dot E_j\dot\gamma_{kl}
    +\frac{1}{2N^2}
    \left(\kappa_3\gamma^{ij}\gamma^{kl}
    +\kappa_4\gamma^{ik}\gamma^{jl}\right)
    \dot E_i M_j\dot\gamma_{kl}
    \notag \\ & \quad 
    +\frac{1}{2N^2}
    \left(\kappa_5\gamma^{ij}\gamma^{kl}
    +\kappa_6\gamma^{ik}\gamma^{jl}\right)
    E_i \dot M_{j}\dot\gamma_{kl}
    +\frac{1}{2N^2}
    \left(\kappa_7\gamma^{ij}\gamma^{kl}
    +\kappa_8\gamma^{ik}\gamma^{jl}\right)
    M_i \dot M_{j}\dot\gamma_{kl}\bigg],\label{eq:lag-dynamical-met}
\end{align}
where $\mathcal{K}^{ij,kl}$ is written in terms of $\gamma^{ij}$, $M^i$, and $E^i$, and the coefficients $\kappa_1,\dots,\kappa_8$ are expressed in terms of $\eta_i$ and $\xi_i$.
For our purpose, we only need the explicit expressions for $\kappa_1,\dots,\kappa_4$:
\begin{align}
    (1+\alpha^2)\kappa_1&=\eta_1+4\xi_1-\xi_2+4\alpha(\eta_3+2\xi_3)
    +\alpha^2(4-\eta_1-4\xi_1+\xi_2),
    \\ 
    (1+\alpha^2)\kappa_2&=-\eta_1-\xi_2-4\alpha\eta_3-\alpha^2(4-\eta_1+\xi_2),
    \\
    \frac{1}{2}(1+\alpha^2)\kappa_3&=\eta_3+2\xi_3
    +\alpha(1-\eta_1-4\xi_1+\xi_2)-\alpha^2(\eta_3+2\xi_3)-\alpha^3,
    \\ 
    \frac{1}{2}(1+\alpha^2)\kappa_4&=-\eta_3
    -\alpha(1-\eta_1)+\alpha^2\eta_3+\alpha^3.
\end{align}
Now, it can be seen from the second line of Eq.~\eqref{eq:lag-dynamical-met} that $E_i$ becomes dynamical in general when gravity is switched on.
We impose $\kappa_1=\kappa_2=\kappa_3=\kappa_4=0$ so that $E_i$ remains nondynamical.
These conditions allow us to express the coefficients in the action in terms of $\alpha$ and $\xi_1$ as follows:
\begin{align}
    &\eta_1=-(1-\alpha^2)\xi_1,
    \quad 
    \eta_2=\alpha^2-2(1-\alpha^2)\xi_1,
    \quad 
    \eta_3=-\alpha(1+2\xi_1),
    \notag \\ & 
    \eta_4=-2\alpha(1+4\xi_1),\quad 
    \xi_2=2(1+\alpha^2)\xi_1,\quad \xi_3=\alpha\xi_1.
\end{align}
We have thus obtained the action characterized by the two parameters $\alpha$ and $\xi_1$.

Let us now define a new variable as
\begin{equation}
    A_{\mu\nu}=B_{\mu\nu}+\alpha\widetilde{B}_{\mu\nu}.
    \label{redefinition of 2-form}
\end{equation}
Interestingly, the resultant action can then be written as
\begin{equation}
    S
    =\int \D^4x\sqrt{-g}\bigg[
    \frac{M_{\text{Pl}}^{2}}{2}R
    -\frac{1}{12}F_{\mu\nu\rho}F^{\mu\nu\rho}
    +\beta A^{\mu\nu}A^{\alpha\beta}L_{\mu\nu\alpha\beta}
    -V(X_A,Y_A)
    \bigg],
    \label{ghost free action 1}
\end{equation}
where $\beta:=\xi_1/2$,
\begin{align}
    F_{\mu\nu\rho}:=3\nabla_{[\mu}A_{\nu\rho]},
    \quad 
    X_A:=A^{\mu\nu}A_{\mu\nu}=\left(1-\alpha^2\right)X_B+2\alpha Y_B,
    \quad 
    Y_A:=A^{\mu\nu}\widetilde{A}_{\mu\nu}=-2\alpha X_B+\left(1-\alpha^2\right)Y_B,
\end{align}
and
\begin{equation}
    L_{\mu\nu\alpha\beta}
    =
    R_{\mu\nu\alpha\beta}
    +R_{\mu\beta}g_{\nu\alpha}
    +R_{\nu\alpha}g_{\mu\beta}
    -R_{\mu\alpha}g_{\nu\beta}
    -R_{\nu\beta}g_{\mu\alpha}
    +\frac{1}{2}R\left(g_{\mu\alpha}g_{\nu\beta}
    -g_{\mu\beta}g_{\nu\alpha}\right)
    \label{double dual Riemann tensor}
\end{equation}
is the double dual Riemann tensor.
Aside from the one in the potential, the Levi-Civita tensor is absorbed into the field redefinition~\eqref{redefinition of 2-form} and disappears from the action.
Although we started from the action containing the dual 2-form field, $\wB_{\mu\nu}$, we have arrived at the kinetic term and the nonminimal coupling to the curvature that are essentially equivalent to those discussed in Ref.~\cite{Heisenberg:2019akx}.
This is one of the main results of the present paper.

\section{Triplet of 2-form fields}

Let us now turn to cosmology in the presence of 2-form fields at the level of the homogeneous and isotropic background.
A single 2-form field is incompatible with isotropy, as in the case of a single vector field.\footnote{Note, however, that Refs.~\cite{Aashish:2018lhv,Aashish:2019zsy,Aashish:2021gdf,Ajith:2025rty}, in which isotropic cosmology with a single 2-form field is considered by imposing an additional constraint to remove the off-diagonal components of the energy-momentum tensor.}
In the case of the vector field, one can circumvent this problem by introducing a triplet of mutually orthogonal vector fields as has been done in Refs.~\cite{Bento:1992wy,Armendariz-Picon:2004say,Germani:2009iq,Emami:2016ldl}.
This type of configuration has also been used for non-Abelian gauge fields~\cite{Maleknejad:2011jw,Adshead:2012kp}.
Here, we introduce a triplet of 2-form fields to ensure compatibility with isotropy in a similar way~\cite{Ajith:2022wia}.

We extend the single 2-form field action~\eqref{ghost free action 1} to accommodate a triplet of 2-form fields with internal O(3) symmetry,
\begin{align}
    S
    =\int \D^4x \sqrt{-g}\left[\frac{M_{\textrm{Pl}}^{2}}{2}R
    -\frac{1}{12}F^{(a)}_{\mu\nu\rho}F^{\mu\nu\rho}_{(a)}
    +\beta A^{\mu\nu}_{(a)}A^{\alpha\beta}_{(a)}L_{\mu\nu\alpha\beta}
    -V(X, Y)+\mathcal{L}_{\textrm{m}}
    \right],
    \label{ghost free triplet action 1}
\end{align}
where $F^{(a)}_{\mu\nu\rho}:=3\partial_{[\mu}A^{(a)}_{\nu\rho]}, X:=A^{\mu\nu}_{(a)}A_{\mu\nu}^{(a)}$, and $Y:=A^{\mu\nu}_{(a)}\widetilde{A}_{\mu\nu}^{(a)}$, with $a=1,2,3$. (Repeated indices $(a)$ are understood to be summed over.)
We have added a matter Lagrangian $\mathcal{L}_{\textrm{m}}$.
The configuration compatible with homogeneous and isotropic cosmology is given by
\begin{align}
    &\D s^2=-N^2(t)\D t^2+a^2(t)\delta_{ij}\D x^i\D x^j,
    \label{eq:FLRW-metric}
    \\ 
    &A^{(a)}_{0i}=-\frac{Na}{\sqrt{6}}
    \phi_{E}(t)\delta^{a}_{i}, \quad
    A^{(a)}_{ij}=\frac{a^2}{\sqrt{6}}
    \epsilon_{ija}\phi_{M}(t).
    \label{triplet anzats}
\end{align}
By computing the energy-momentum tensor $T_{\mu\nu}$ for the 2-form fields explicitly, one can see that the off-diagonal components indeed vanish.

Substituting Eqs.~\eqref{eq:FLRW-metric} and~\eqref{triplet anzats} into the action~\eqref{ghost free triplet action 1}, we obtain
\begin{align}
    S=
   \int \D t \D^3x\, Na^3\left[
   -\frac{3M_\textrm{Pl}^2 {\dot{a}}^2}{N^2 a^2}
   +\frac{\dot{\phi}_{M}^2}{4N^2}
   -\frac{2\beta\phi_{E}^2 {\dot{a}}^2}{N^2 a^2}
   +\frac{(1-4\beta)\phi_{M}\dot{\phi}_{M}\dot{a}}{N^2 a}
   +\frac{(1-4\beta)\phi_{M}^2 {\dot{a}}^2}{N^2 a^2}
   -V(X, Y)+\mathcal{L}_{\textrm{m}}
   \right],
   \label{ghost free triplet action 2}
\end{align}
with
\begin{align}
    X=-\phi_E^2+\phi_M^2,
    \quad 
    Y=2\phi_E\phi_M.
\end{align}
Varying the above action with respect to $N$, $a$, and $\phi_M$, we obtain, respectively,
\begin{align}
    3M_\textrm{Pl}^2 H^2
    &=\rho+
    \frac{1}{4}\dot{\phi}_{M}^2
    +V
    +(1-4\beta)H\phi_{M}\dot{\phi}_{M}
    +(1-4\beta)H^2\phi_{M}^2
    -2\beta H^2\phi_{E}^2,
    \label{triplet eom N}
    \\
    M_\textrm{Pl}\left(3 H^2+2\dot{H}\right)
    &=-P+
    \frac{1}{12}(1-16\beta)\dot{\phi}_{M}^2
    +\frac{1}{3}(1-4\beta)\phi_{M}\Ddot{\phi}_{M}
    -2\beta H^2\phi_{E}^2
    +(1-4\beta)H^2\phi_{M}^2
    \notag \\
    &\quad 
    -\frac{4}{3}\beta\phi_{E}^2\dot{H}
    +\frac{2}{3}(1-4\beta)\phi_{M}^2\dot{H}
    -\frac{8}{3}\beta H\phi_{E}\dot{\phi}_{E}
    +\frac{4}{3}(1-4\beta)H\phi_{M}\dot{\phi}_{M}+V,
    \label{triplet eom a}
\end{align}
and 
\begin{align}
    \Ddot{\phi}_{M}
    +3H\dot{\phi}_{M}
    +2(1-4\beta)H^2\phi_{M}
    +2(1-4\beta)\phi_{M}\dot{H}
    +4V_X\phi_M
    +4V_Y\phi_E
    =0,
    \label{triplet eom M}
\end{align}
where $\rho$ and $P$ are, respectively, the energy density and pressure of the matter fields in $\mathcal{L}_{\textrm{m}}$, $H:=\dot a/a$, and we set $N=1$.
Varying the action with respect to $\phi_E$ gives the constraint equation,
\begin{align}
    (2\beta H^2-V_X) \phi_{E}
    +\phi_M V_Y
    =0,
    \label{triplet constraint E}
\end{align}
where $V_X=\partial V/\partial X$ and $V_Y=\partial V/\partial Y$.

The constraint equation~\eqref{triplet constraint E} can (in principle) be solved for $\phi_E$, and hence one can remove $\phi_E$ from Eqs.~\eqref{triplet eom N}--\eqref{triplet eom M}.
The resultant system is somewhat similar to a cosmological background with a nonminimally coupled scalar field.
An important remark here is that if the potential depends only on $X$, i.e., $V_Y=0$, Eq.~\eqref{triplet constraint E} implies that $\phi_E=0$. However, in general, both $\phi_M$ and $\phi_E$ are nonvanishing.

\section{Tensor perturbations}

An interesting consequence of introducing a triplet of 2-form fields is that additional tensor modes emerge on top of the usual tensor-type metric perturbations.\footnote{Tensor modes arising from the triplet of the 2-form fields have not been taken into account in Ref.~\cite{Ajith:2022wia}.}
Furthermore, under certain conditions, gravitational parity violation can occur in our setup.
In this section, we study tensor perturbations on the cosmological background introduced above, in order to clarify how the interesting phenomenology arises.

\subsection{Quadratic action and stability}
\label{Quadratic action and stability}

We first derive the quadratic action for tensor perturbations,
which gives us stability conditions as well as equations of motion.

The metric with the transverse and traceless tensor perturbations $h_{ij}$ is given by
\begin{align}
    \D s^2=-\D t^2+a^2(e^h)_{ij}\D x^i\D x^j,
    \label{metric tensor perturbation}
\end{align}
where $(e^h)_{ij}=\delta_{ij}+h_{ij}+(1/2)h_{ik}h_{kj}+\dots$.
Tensor perturbations also arise from the 2-form field as
\begin{align}
    A_{0i}^{(a)}&=-\frac{a}{\sqrt{6}}\left(
    \phi_E\delta_i^a+E_{ia}
    \right),
    \\
    A_{ij}^{(a)}&=\frac{a^2}{\sqrt{6}}\epsilon_{ijk}\left(
    \phi_M\delta_k^a+M_{ka}
    \right),
    \label{2-form tensor perturbation}
\end{align}
where $E_{ij}$ and $M_{ij}$ are tranverse and traceless perturbations.
Indices of $h_{ij}, E_{ij}$, and $M_{ij}$ are raised and lowered with $\delta_{ij}$.
We assume that the matter Lagrangian gives rise to no tensor-type perturbations.

We expand the action to second order in the tensor perturbations and move to Fourier space to get
\begin{align}
    S
    &=\sum_{A=\textrm{R,L}}\frac{1}{(2\pi)^3}\int \D t\D^3k \, a^3
    \biggl[a_1 \dot{h}^{A}\dot{\Bar{h}}^{A}
    +a_2 \dot{h}^{A}\dot{\Bar{M}}^{A}
    +a_3 \dot{M}^{A}\dot{\Bar{M}}^{A}
    +b_1 E^{A}\dot{\Bar{h}}^{A}
    +b_2 M^{A}\dot{\Bar{h}}^{A}
    +b_3 E^{A}\dot{\Bar{M}}^{A}
    \notag \\
    &\quad+c_1 h^{A}\bar{h}^{A}
    +c_2 h^{A}\bar{E}^{A}
    +c_3 h^{A}\bar{M}^{A}
    +c_4 E^{A}\bar{E}^{A}
    +c_5 E^{A}\bar{M}^{A}
    +c_6 M^{A}\bar{M}^{A}
    +(\text{c.c.})
    \biggr],
    \label{ghost free triplet tensor action  1}
\end{align}
where complex conjugation is indicated by a bar,
and the explicit expressions for the coefficients $a_1, \dots, c_6$ are presented in Appendix~\ref{app:cfs}.
Our conventions here are as follows.
The Fourier transform of a tensor perturbation $T_{ij}(t,\mathbf{x})$ is given by
\begin{equation}
    T_{ij}(t,\mathbf{x})=\sum_{A=\textrm{R,L}}\frac{1}{(2\pi)^3}\int \D^3 k\, T^{A}(t,\mathbf{k})e^{A}_{ij}(\hat{\mathbf{k}})e^{i\mathbf{k}\cdot\mathbf{x}},
    \label{Fourier trans for tensor}
\end{equation}
where $e^A_{ij}(\hat{\mathbf{k}})$ ($A=\,$R, L) is the circular polarization basis with $\hat{\mathbf{k}}:=\mathbf{k}/|\mathbf{k}|$.
The normalization of the circular polarization basis is given by $e^{A}_{ij}\Bar{e}^{A'}_{ij}=2\delta^{AA'}$. We use the relation
\begin{align}
    \epsilon_{ilk}\hat k_le_{jk}^{A}=-i\lambda^{A}e_{ij}^A,
    \label{parity circular basis}
\end{align}
where $\lambda^{\textrm{R}}=+1$ and $\lambda^{\textrm{L}}=-1$.

Since there is no time derivative acting on $E^A$, it can be removed from the action by the use of the constraint equation obtained by variation with respect to $\Bar{E}^{A}$:
\begin{align}
    b_1 \dot{h}^{A}+b_3 \dot{M}^{A}+c_2 h^{A}+c_4E^{A}+ c_5 M^{A}=0
    .
\end{align}
The reduced action (for each polarization state) expressed in terms of the dynamical variables $\psi^1:=M_{\textrm{Pl}} h$ and $\psi^2:=M$ is given by
\begin{align}
    S^{\text{R,L}}= \int\D t\D^3k \,a^3\left[
    \mathcal{A}_{IJ}\dot{\bar\psi}^{I}\dot{\psi}^J
    -\mathcal{B}_{IJ}\left(\bar\psi^I\dot\psi^J+\psi^I\dot{\bar\psi}^J\right)
    -\mathcal{C}_{IJ}\bar\psi^I\psi^J
    \right],
    \label{action for tensor pert Fourier 1}
\end{align}
where the polarization label $A$ has been omitted.
The coefficients $\mathcal{A}_{IJ}$ and $\mathcal{C}_{IJ}$ are symmetric, while $\mathcal{B}_{IJ}$ is antisymmetric, with respect to $I$ and $J$.
Explicitly, we have 
\begin{align}
    \mathcal{A}_{11}&=\frac{1}{3\mpl^2(U+k^2/4a^2)}
    \biggl\{
    \frac{1}{4}\left[
        -2\beta H^2\left(3\mpl^2+6\beta\phi_E^2-4\beta\phi_M^2\right)
        +V_X\left(3\mpl^2+2\beta\phi_E^2-4\beta\phi_M^2\right)
    \right]
    \notag\\ & \quad 
        +\frac{4\lambda^Ak}{a}\beta^2H\phi_E\phi_M
        +\frac{k^2}{16 a^2}\left[3\mpl^2+2\beta\phi_E^2-4\beta(1+8\beta)\phi_M^2\right]
    \biggr\},
    \label{eq. A_11}
    \\
    \mathcal{A}_{12}&=-\frac{2\beta\left[U\phi_M+(\lambda^Ak/4a)H\phi_E\right]}{3\mpl(U+k^2/4a^2)},
    \label{eq. A_12}
    \\
    \mathcal{A}_{22}&=\frac{1}{6}\frac{U}{U+k^2/4a^2},
    \label{eq. A_22}
    \\ 
    \mathcal{B}_{12}&=\frac{1}{12\mpl (U+k^2/4a^2)}\biggl\{
        \frac{k^2}{a^2}\beta(\dot \phi_M-H\phi_M)+
        \frac{\lambda^Ak}{a}\left(
            2\beta H^2-U
        \right)\phi_E
        \notag \\ & \quad 
        +\left[(1+4\beta)\dot\phi_M+2(1+2\beta)H\phi_M\right]
        -2\beta H\left[
            (1+4\beta)H\dot\phi_M+2(1+2\beta)H^2\phi_M-2\phi_EV_Y
        \right]
    \biggr\},
    \label{eq:def:B12}
\end{align}
with
\begin{equation}
    U:=V_{X}-2\beta H^2 ,
    \label{def of U}
\end{equation}
while the explicit expression for $\mathcal{C}_{IJ}$ is messy.

Let us first investigate the structure of the kinetic matrix $\mathcal{A}_{IJ}$.
To avoid the ghost instability, we require that
\begin{align}
    \det \mathcal{A}&=
    \frac{3\mpl^2 U-2\beta (4\beta H^2-U)\phi_{E}^2
    -4\beta(1+8\beta)U\phi_{M}^2}{72\mpl^2(U+k^2/4a^2)}
    >0
    \label{ghost free condition for tensor perturbations 1}
\end{align}
and $\mathcal{A}_{22}>0$.
Assuming for simplicity that
$|\beta|^{1/2}\phi_{E}, |\beta|^{1/2}\phi_{M}\ll M_{\textrm{Pl}}$,
the ghost-free conditions are satisfied if
\begin{equation}
    U>0.
    \label{ghost free condition for tensor perturbations 2}
\end{equation}
In the minimally-coupled theory, the condition simply reads $V_X>0$.

Next, we investigate the high-$k$ limit of the coefficients $\mathcal{A}_{IJ}$, $\mathcal{B}_{IJ}$, and $\mathcal{C}_{IJ}$ to see the propagation speeds of the two tensor modes.
In the high-$k$ limit, we have
\begin{align}
    \mathcal{A}_{11}&\approx
    \frac{3\mpl^2+2\beta\phi_E^2-4\beta(1+8\beta)\phi_M^2}{12\mpl^2},
    \\
    \mathcal{A}_{12}&\approx-\frac{2\beta}{3\mpl}H\phi_E\left(\frac{\lambda_Ak}{a}\right)^{-1}
    ,
    \\ 
    \mathcal{A}_{22}&\approx\frac{2U}{3}\left(\frac{k}{a}\right)^{-2},
    \\ 
    \mathcal{B}_{12}&\approx\frac{\beta}{3M_{\textrm{Pl}}}\left(\dot\phi_M-H\phi_M\right),
    \label{eq. B_12}
    \\ 
    \mathcal{C}_{11}&\approx\frac{3\mpl^2+2\beta(3+16\beta)\phi_E^2}{12\mpl^2}\frac{k^2}{a^2}
    ,
    \label{eq. C_11}
    \\ 
    \mathcal{C}_{12}&\approx-\frac{2\beta\dot\phi_E}{3\mpl}\frac{\lambda_A k}{a},
    \label{eq. C_12}
    \\
    \mathcal{C}_{22}&\approx\frac{2}{3}\left(U-2\beta\dot H\right).
    \label{eq. C_22}
\end{align}
Assuming the solution of the form $\psi^I\sim \psi_0^I\exp[-ic_{\textrm{T}}k\int^t\D t'/a(t')]$, the field equations in the high-$k$ limit yields
\begin{align}
    \mathcal{D}_{IJ}\psi^{J}_0=0,
    \label{propergating speed instability 1}
\end{align}
with
\begin{align}
    \mathcal{D}_{IJ}=
    \frac{c_\mathrm{T}^{2}k^2}{a^2}\mathcal{A}_{IJ}+2i\frac{c_\mathrm{T}k}{a}\mathcal{B}_{IJ}-\mathcal{C}_{IJ}.
\end{align}
We have
\begin{align}
    \textrm{det}(\mathcal{D}_{IJ})=\frac{k^2}{6a^2}
    \left[
    d_2(c^2_{\mathrm{T}})^2-2d_1c^2_{\mathrm{T}}+d_0
    \right]+\mathcal{O}(k),
\end{align}
where
\begin{align}
    d_0&=
     U-2\beta \dot{H}-\frac{2\beta}{3\mpl^2}\left[
    -(3+16\beta)(U-2\beta \dot H)\phi_E^2
    +4\beta\dot{\phi}_E^2
    \right],
    \\ 
    d_1&=
    U-\beta \dot{H}
    -\frac{2\beta}{3\mpl^2}\biggl\{
    U\left[ -2(1+4\beta)\phi_E^2+(1+8\beta)\phi_M^2 \right]
    \notag \\
    &\quad
    +\beta\left[
    -2H^2\phi_M^2+\phi_E^2\dot{H}+4H(\phi_E\dot{\phi}_E+\phi_M\dot{\phi}_M)
    -2
    (1+8\beta)\phi_M^2\dot{H}-2\dot{\phi}_M^2
    \right]
    \biggr\},
    \\ 
    d_2&=
    U-\frac{2\beta}{3\mpl^2}\left\{
    4\beta H^2\phi_E^2+U\left[
    -\phi_E^2+2(1+8\beta)\phi_M^2
    \right]
    \right\},
\end{align}
and hence the propagation speeds can be obtained by solving
$d_2(c^2_{\mathrm{T}})^2-2d_1 c^2_{\mathrm{T}}+d_0=0$.
Assuming again $|\beta|^{1/2}\phi_{E}, |\beta|^{1/2}\phi_{M}\ll M_{\textrm{Pl}}$ for simplicity, we find
\begin{align}
    c^2_{\mathrm{T}}=1,\quad 1-\frac{2\beta\dot H}{U}.
\end{align}
It can be seen that one of the propagation speeds deviates from the speed of light if $\beta\neq 0$, and the gradient instability can be avoided if $U>2\beta\dot H$.

\subsection{Chiral gravitational waves}

We are now in a position to discuss the conditions under which chiral gravitational waves can be produced.
In the reduced action~\eqref{action for tensor pert Fourier 1}, the terms with $\lambda^A$ cause gravitational parity violation, though they are hidden in the coefficients $\mathcal{A}_{IJ}$, $\mathcal{B}_{IJ}$, and $\mathcal{C}_{IJ}$, and are not shown explicitly.
From Eqs.~\eqref{eq. A_11}--\eqref{eq:def:B12}, we see that the $\lambda^A$-dependent terms in $\mathcal{A}_{IJ}$ and $\mathcal{B}_{IJ}$ are proportional to $\phi_E$.
Although the explicit expressions are messy, one can also check that all the $\lambda^A$-dependent terms in $\mathcal{C}_{IJ}$ vanish if $\phi_E=V_Y=V_{XY}=0$.
Recalling that the constraint equation~\eqref{triplet constraint E} implies $\phi_E=0$ if $V_Y=0$, we conclude that the crucial ingredient for chiral gravitational waves is that the potential depends on $Y$ and $\phi_E\neq 0$.

\section{Conclusions}\label{sec:conclusions}

In this paper, we have studied gravitational parity violation in ghost-free 2-form field theories.
We have started from the action involving all possible kinetic terms and dimension four operators that couple the curvature tensors with a 2-form field $B_{\mu\nu}$ and its dual $\wB_{\mu\nu}$~\cite{Altschul:2009ae}.
The coupling to $\wB_{\mu\nu}$ is relevant for gravitational parity violation~\cite{Manton:2024hyc}.
We carefully revisited identities and expressions that reduce to total derivatives, and identified the term that was considered in the literature but is in fact redundant.
Several relations among the parameters in the action have been imposed so that no additional propagating degrees of freedom arise relative to the massive 2-form field theory.
The analysis has been done by examining the general kinetic terms in Minkowski spacetime and the nonminimal couplings in homogeneous cosmological models.
Although we thus obtained a two-parameter family of healthy 2-form field theories, we found that the dual 2-form in the kinetic terms and nonminimal couplings can be absorbed into a field redefinition of the form $B_{\mu\nu}\to A_{\mu\nu}=B_{\mu\nu}+\alpha \wB_{\mu\nu}$, leaving in the end a one-parameter family of theories, which has essentialy the same action as that proposed by Heisenberg and Trenkler~\cite{Heisenberg:2019akx}.
The dual of the new field, $\widetilde{A}_{\mu\nu}$, appears only in the potential.

Since anisotropic components of the energy-momentum tensor are generated by a single 2-form field in general, we need three copies of 2-form fields, $A_{\mu\nu}^{(a)}$ $(a=1,2,3)$, to implement them in a homogeneous and isotropic universe, as has been done in the case of vector fields.
The introduction of the triplet of 2-form fields enables us to study the isotropic cosmological background and tensor perturbations consistently.
In this setup, there are three types of tensor perturbations: one originating from the metric and two from the 2-form fields.
One of the latter two is an auxiliary variable and can be eliminated by the use of the constraint equation.
We have obtained the quadratic action for the tensor perturbations describing the coupled system of the two dynamical variables.
By inspecting this quadratic action, we have shown that the necessary conditions for gravitational parity violation to occur are that the potential depends on $\widetilde{A}_{\mu\nu}^{(a)}$ and that the background ``electric'' components (i.e., $A_{0i}^{(a)}$) do not vanish.

This paper focused solely on the construction of healthy 2-form field theories nonminimally coupled to gravity and the derivation of the formal conditions for parity violation in gravitational waves.
It would be important to develop concrete models, e.g., for inflation in the presence of the triplet of 2-form fields and explore their implications for observations.
Since next-generation missions for gravitational-wave observations would be able to offer opportunities for testing the chirality of gravitational waves, this is an important direction for future work.

%--- Acknowledgments ---%--- Acknowledgments ---%--- Acknowledgments ---%
\acknowledgments
We would like to thank Anamaria Hell,
Ippei Obata, and Daisuke Yamauchi for interesting discussions.
The work of YH was supported by
the Rikkyo University Special Fund for Research and Asahi Glass Foundation Scholarship Program.
The work of TM was supported by Specific Project Grant from TMCIT.
The work of TK was supported by
JSPS KAKENHI Grant No.~JP25K07308 and
MEXT-JSPS Grant-in-Aid for Transformative Research Areas (A) ``Extreme Universe'',
No.~JP21H05182 and No.~JP21H05189.
%--- Acknowledgments ---%--- Acknowledgments ---%--- Acknowledgments ---%

\appendix

\section{The case of a nonminimally coupled vector field}\label{app:vector}

It is instructive to see how one can constrain the form of the kinetic terms and the coupling to gravity of a vector field $A_\mu$ (see e.g.~\cite{Esposito-Farese:2009wbc}).
We start with the following Lagrangian of a vector field coupled nonminimally to gravity,
\begin{align}
    \mathcal{L}=\frac{\mpl^2}{2}R-\frac{1}{4}F_{\mu\nu}F^{\mu\nu}+\frac{\eta_1}{2}
    \left(\nabla_\mu A^\mu\right)^2
    +\frac{\eta_2}{2}\nabla_\mu A_\nu \nabla^\nu A^\mu-V(X)+
    \frac{\xi_1}{2}XR+\frac{\xi_2}{2}A^\mu A^\nu R_{\mu\nu},
\end{align}
where $F_{\mu\nu}:=\nabla_\mu A_\nu-\nabla_\nu A_\mu$ and $X:=A_\mu A^\mu$.
Here, $\eta_1$, $\eta_2$, $\xi_1$, and $\xi_2$ are constant parameters.
This includes all possible kinetic terms and dimension four coupling to the curvature tensors.
Note the relation
\begin{align}
    \nabla_\mu\left(A^\mu\nabla_\nu A^\nu -A^\nu\nabla_\nu A^\mu \right)
    =\left(\nabla_\mu A^\mu\right)^2-\nabla_\mu A_\nu \nabla^\nu A^\mu-A^\mu A^\nu R_{\mu\nu},
\end{align}
which allows us to set $\eta_2=0$ without loss of generality.

\subsection{Vector field in Minkowski}

Let us first switch off gravity and consider the dynamics of the vector field in a fixed Minkowski background. We write $A_0=-\mathcal{A}$ and $A_i=\partial_iA^{\textrm{L}}+A_i^{\textrm{T}}$,
where $A^{\textrm{L}}$ is the longitudinal mode and $A_i^{\textrm{T}}$ is the transverse mode
satisfying $\partial^i A_i^{\textrm{T}}=0$.
The Lagrangian is then written as
\begin{align}
    \mathcal{L}&=
    \mathcal{L}_{\textrm{kin}}^{\textrm{L}}
    +\frac{1}{2}\left[
    \dot A^{\textrm{T}}_i\dot A^{\textrm{T}}_i-A^{\textrm{T}}_i(-\Delta)A^{\textrm{T}}_i
    \right]-V(X),\label{eq:ATLLag}
    \\ 
    \mathcal{L}_{\textrm{kin}}^{\textrm{L}}&=
    \frac{1}{2}\left(\dot{A}^{\textrm{L}}+\mathcal{A}\right)(-\Delta)
    \left(\dot{A}^{\textrm{L}}+\mathcal{A}\right)+
    \frac{\eta_1}{2}\left(\dot{\mathcal{A}}+\Delta A^{\textrm{L}}\right)^2,
\end{align}
where $\Delta=\partial_i\partial^i$.
It is obvious that the transverse sector is healthy.
To see that instabilities show up when $\eta_1\neq 0$,
let us rewrite the kinetic part of the Lagrangian $\mathcal{L}_{\textrm{kin}}^{\textrm{L}}$
by introducing an auxiliary variable as
\begin{align}
    \mathcal{L}_{\textrm{kin}}^{\textrm{L}}=
    -\frac{1}{2}\chi(-\Delta)\chi+\left(\dot{A}^{\textrm{L}}
    +\mathcal{A}\right)(-\Delta)\chi 
    +\frac{\eta_1}{2}\left(\dot{\mathcal{A}}+\Delta A^{\textrm{L}}\right)^2.
\end{align}
For simplicity, let us ignore for the moment the potential term.
The equation of motion for $A^{\textrm{L}}$ yields the constraint
\begin{align}
    \dot\chi+ \eta_1\left(\dot{\mathcal{A}}+\Delta A^{\textrm{L}}\right)=0.
\end{align}
Using this, one can remove $A^{\textrm{L}}$ from $\mathcal{L}_{\textrm{kin}}^{\textrm{L}}$ to get
\begin{align}
    \mathcal{L}^{\textrm{L}}_{\textrm{kin}}=-\frac{1}{2\eta_1}\dot\chi^2-\dot\chi\dot{\mathcal{A}}
    -\frac{1}{2}\chi(-\Delta)\chi+\mathcal{A}(-\Delta)\chi.
\end{align}
Peforming the change of variables $\psi=\chi+\eta_1\mathcal{A}$,
we have
\begin{align}
    \mathcal{L}_{\textrm{kin}}^{\textrm{L}}=
    \frac{\eta_1}{2}\dot{\mathcal{A}}^2
    -\frac{1}{2\eta_1}\dot\psi^2
    +\frac{1}{2}\mathcal{A}(-\Delta)\mathcal{A}
    -\frac{1}{2}\left[\psi-(1+\eta_1)\mathcal{A}\right](-\Delta)
    \left[\psi-(1+\eta_1)\mathcal{A}\right].
\end{align}
The signs of the two kinetic terms are always opposite,
indicating that one of the two fields $\psi$ and $\mathcal{A}$ is a ghost.
This leads us to focus on the case of $\eta_1=0$.

In the case of $\eta_1=0$, Eq.~\eqref{eq:ATLLag} reads 
\begin{align}
    \mathcal{L}=\frac{1}{2}\left(\mathcal{A}+\dot{A}^{\textrm{L}}\right)
    (-\Delta)\left(\mathcal{A}+\dot{A}^{\textrm{L}}\right)
    +\frac{1}{2}\left[
    \dot A^{\textrm{T}}_i\dot A^{\textrm{T}}_i-A^{\textrm{T}}_i(-\Delta)A^{\textrm{T}}_i
    \right]-V(X).
\end{align}
The equation of motion for $\mathcal{A}$ yields the constraint 
\begin{align}
    -\Delta\left(\mathcal{A}+\dot{A}^{\textrm{L}}\right)+
    2\mathcal{A}V_{,X}=0,
\end{align}
using which one can remove $\mathcal{A}$ from the action.
Thus, in the case of $\eta_1=0$,
we are left with the two dynamical variables $A^{\textrm{L}}$ and $A^{\textrm{T}}$
(as long as $V_X\neq 0$).

\subsection{Coupling to gravity}

In the above analysis in the fixed Minkowski background,
one cannot extract any information on $\xi_1$ and $\xi_2$.
However, once gravity is switched on, mixing with gravity would render $\mathcal{A}$ dynamical.
To see how this occurs, we consider a homogeneous vector field in a spatially flat, homogeneous background,
\begin{align}
    A_0&=-N\mathcal{A}(t), \quad A_i=A_i(t),
    \\ 
    \D s^2&=-N^2(t)\D t^2+\gamma_{ij}(t)\D x^i\D x^j.
\end{align}
Substituting this ansatz into the Lagrangian, we get 
\begin{align}
        \mathcal{L}=\frac{1}{2N^2}\mathcal{K}^{ij,kl}\dot\gamma_{ij}\dot\gamma_{kl}-V 
        +\frac{\eta_1}{2N^2}\dot{\mathcal{A}}^2 
        +\frac{\eta_1+2\xi_1+\xi_2}{2N^2}\gamma^{ij}\mathcal{A}\dot\gamma_{ij}\dot{\mathcal{A}}
        +\frac{1}{2N^2}\gamma^{ij}\dot A_i\dot A_j 
        %\notag \\ &\quad 
        -\frac{1}{2N^2}\left(2\xi_1A^k\gamma^{ij}+\xi_2A^i\gamma^{jk}\right) 
        \dot\gamma_{ij}\dot{A}_k,
\end{align}
where $\mathcal{K}^{ij,kl}$ is written in terms of $\gamma_{ij}$, $\mathcal{A}$, and $A_i$, but it does not contain their time derivatives.
From the previous analysis we set $\eta_1=0$.
One further notices that the term $\dot \gamma_{ij}\dot{\mathcal{A}}$ gives rise to a second time derivative of $\gamma_{ij}$ in the equation of motion for $\mathcal{A}$ unless
\begin{align}
    2\xi_1+\xi_2=0.
\end{align}
When this condition is met, we have
\begin{align}
    \frac{\xi_1}{2}XR+\frac{\xi_2}{2}A^\mu A^\nu R_{\mu\nu}
    =\frac{\xi_2}{2}A^\mu A^\nu G_{\mu\nu}.
\end{align}
This is precisely the nonminimal coupling found in the generalized Proca theory~\cite{Heisenberg:2014rta}, which has been known to be ghost-free.
One can thus single out the healthy nonminimal coupling of a vector field with gravity by analyzing an inhomogeneous configuration of the vector field in a Minkowski background and a homogeneous field in a spatially flat, homogeneous background.
In the main text, we study the case of a 2-form field in the same way as presented in this appendix.

\section{Coefficients in the quadratic action~\eqref{ghost free triplet tensor action  1}}
\label{app:cfs}

The coefficients in the action~\eqref{ghost free triplet tensor action  1} are given by
\begin{align}
    a_1&=
    \frac{1}{24}\left(3M_{\textrm{Pl}}^2+2\beta \phi_{E}^2-4\beta\phi_{M}^2\right),
    \label{coefficient: a1}
    \\
    a_2&=-\frac{2}{3}\beta\phi_{M},
    \label{coefficient: a2}
    \\
    a_3&=\frac{1}{12},
    \label{coefficient: a3}
    \\
    b_1&=\frac{2\beta}{3}\left(-\frac{\lambda^{A}k}{a}\phi_{M}+H\phi_{E}\right),
    \label{coefficient: b1}
    \\
    b_2&=\frac{1}{6}\left[
    -\frac{4\lambda^{A}k}{a}\beta\phi_{E}
    -2(1+2\beta)H\phi_{M}
    -(1+4\beta)\dot{\phi}_{M}
    \right],
    \label{coefficient: b2}
    \\
    b_3&=\frac{\lambda^{A}k}{6a},
    \label{coefficient: b3}
    \\
    c_1&=\frac{1}{24}\bigg\{
    -3\frac{k^2}{a^2}\left(M_{\textrm{Pl}}^2+2\beta\phi_{E}^2\right)
    +4H^2\left[\beta\phi_{E}^2+(1+5\beta)\phi_{M}^2\right]
    +12\beta\phi_{M}^2\dot{H}
    +(1+8\beta)\dot{\phi}_{M}^2
    \notag\\
    &\quad
    +4H\left[2\beta\phi_{E}\dot{\phi}_{E}+(1+8\beta)\phi_{M}\dot{\phi}_{M}\right]
    +8\beta\phi_{M}\Ddot{\phi}_{M}
    -4\phi_{M}^2V_{X}
    +4\phi_{E}^2(\beta \dot{H}+V_{X})
    \bigg\},
    \label{coefficient: c1}
    \\
    c_2&=\frac{1}{6}\left[
    \frac{4k^2}{a^2}\beta\phi_{E}
    +\frac{\lambda^{A}k}{a}(2H\phi_{M}+\dot{\phi}_{M})
    +4\phi_{E}(2\beta H^2-V_X)
    \right],
    \label{coefficient: c2}
    \\
    c_3&=\frac{1}{6}\left\{
    2(-1+4\beta)H^2\phi_{M}-3H\dot{\phi}_{M}-\Ddot{\phi}_{M}
    +2\phi_{M}\left[ (-1+4\beta)\dot{H}-2V_X \right]
    \right\},
    \label{coefficient: c3}
    \\
    c_4&=\frac{1}{12}\left[
    \frac{k^2}{a^2}
    +4(-2\beta H^2+V_X)
    \right],
    \label{coefficient: c4}
    \\
    c_5&=\frac{1}{3}\left(
    \frac{\lambda^{A}k}{a}H
    +2V_Y
    \right),
    \label{coefficient: c5}
    \\
    c_6&=\frac{1}{6}\left[
    (-1+4\beta)(H^2+\dot{H})-2V_X
    \right].
    \label{coefficient: c6}
\end{align}

%-----------------------------------------------------------------%
\bibliography{refs}

@article{Manton:2024hyc,
    author = "Manton, Tucker and Alexander, Stephon",
    title = "{Kalb-Ramond field and gravitational parity violation}",
    eprint = "2401.14452",
    archivePrefix = "arXiv",
    primaryClass = "gr-qc",
    doi = "10.1103/PhysRevD.110.044067",
    journal = "Phys. Rev. D",
    volume = "110",
    number = "4",
    pages = "044067",
    year = "2024"
}

@article{Takahashi:2009wc,
    author = "Takahashi, Tomohiro and Soda, Jiro",
    title = "{Chiral Primordial Gravitational Waves from a Lifshitz Point}",
    eprint = "0904.0554",
    archivePrefix = "arXiv",
    primaryClass = "hep-th",
    doi = "10.1103/PhysRevLett.102.231301",
    journal = "Phys. Rev. Lett.",
    volume = "102",
    pages = "231301",
    year = "2009"
}

@article{Zhu:2013fja,
    author = "Zhu, Tao and Zhao, Wen and Huang, Yongqing and Wang, Anzhong and Wu, Qiang",
    title = "{Effects of parity violation on non-gaussianity of primordial gravitational waves in Ho{\v{r}}ava-Lifshitz gravity}",
    eprint = "1305.0600",
    archivePrefix = "arXiv",
    primaryClass = "hep-th",
    doi = "10.1103/PhysRevD.88.063508",
    journal = "Phys. Rev. D",
    volume = "88",
    pages = "063508",
    year = "2013"
}

@article{Maleknejad:2016qjz,
    author = "Maleknejad, Azadeh",
    title = "{Axion Inflation with an SU(2) Gauge Field: Detectable Chiral Gravity Waves}",
    eprint = "1604.03327",
    archivePrefix = "arXiv",
    primaryClass = "hep-ph",
    doi = "10.1007/JHEP07(2016)104",
    journal = "JHEP",
    volume = "07",
    pages = "104",
    year = "2016"
}

@article{Murata:2024urv,
    author = "Murata, Tomoaki and Kobayashi, Tsutomu",
    title = "{Chromo-natural inflation supported by enhanced friction from Horndeski gravity}",
    eprint = "2408.01773",
    archivePrefix = "arXiv",
    primaryClass = "astro-ph.CO",
    reportNumber = "RUP-24-11",
    doi = "10.1088/1475-7516/2024/10/044",
    journal = "JCAP",
    volume = "10",
    pages = "044",
    year = "2024"
}

@article{Dimastrogiovanni:2023oid,
    author = "Dimastrogiovanni, Ema and Fasiello, Matteo and Michelotti, Martino and Pinol, Lucas",
    title = "{Primordial gravitational waves in non-minimally coupled chromo-natural inflation}",
    eprint = "2303.10718",
    archivePrefix = "arXiv",
    primaryClass = "astro-ph.CO",
    doi = "10.1088/1475-7516/2024/02/039",
    journal = "JCAP",
    volume = "02",
    pages = "039",
    year = "2024"
}

@article{Watanabe:2020ctz,
    author = "Watanabe, Yuki and Komatsu, Eiichiro",
    title = "{Gravitational Wave from Axion-SU(2) Gauge Fields: Effective Field Theory for Kinetically Driven Inflation}",
    eprint = "2004.04350",
    archivePrefix = "arXiv",
    primaryClass = "hep-th",
    month = "4",
    year = "2020"
}

@article{Iarygina:2021bxq,
    author = "Iarygina, Oksana and Sfakianakis, Evangelos I.",
    title = "{Gravitational waves from spectator Gauge-flation}",
    eprint = "2105.06972",
    archivePrefix = "arXiv",
    primaryClass = "hep-th",
    doi = "10.1088/1475-7516/2021/11/023",
    journal = "JCAP",
    volume = "11",
    number = "11",
    pages = "023",
    year = "2021"
}

@article{Bielefeld:2015daa,
    author = "Bielefeld, Jannis and Caldwell, Robert R.",
    title = "{Cosmological consequences of classical flavor-space locked gauge field radiation}",
    eprint = "1503.05222",
    archivePrefix = "arXiv",
    primaryClass = "gr-qc",
    doi = "10.1103/PhysRevD.91.124004",
    journal = "Phys. Rev. D",
    volume = "91",
    number = "12",
    pages = "124004",
    year = "2015"
}

@article{Tishue:2021blv,
    author = "Tishue, Avery J. and Caldwell, Robert R.",
    title = "{Relic cosmological vector fields and inflationary gravitational waves}",
    eprint = "2105.08073",
    archivePrefix = "arXiv",
    primaryClass = "astro-ph.CO",
    doi = "10.1103/PhysRevD.104.063531",
    journal = "Phys. Rev. D",
    volume = "104",
    number = "6",
    pages = "063531",
    year = "2021"
}

@article{Dimastrogiovanni:2016fuu,
    author = "Dimastrogiovanni, Emanuela and Fasiello, Matteo and Fujita, Tomohiro",
    title = "{Primordial Gravitational Waves from Axion-Gauge Fields Dynamics}",
    eprint = "1608.04216",
    archivePrefix = "arXiv",
    primaryClass = "astro-ph.CO",
    doi = "10.1088/1475-7516/2017/01/019",
    journal = "JCAP",
    volume = "01",
    pages = "019",
    year = "2017"
}

@article{Sulantay:2022sag,
    author = "Sulantay, Felipe and Lagos, Macarena and Ba{\~n}ados, M{\'a}ximo",
    title = "{Chiral gravitational waves in Palatini-Chern-Simons gravity}",
    eprint = "2211.08925",
    archivePrefix = "arXiv",
    primaryClass = "gr-qc",
    doi = "10.1103/PhysRevD.107.104025",
    journal = "Phys. Rev. D",
    volume = "107",
    number = "10",
    pages = "104025",
    year = "2023"
}

@article{Capanelli:2023uwv,
    author = "Capanelli, Christian and Jenks, Leah and Kolb, Edward W. and McDonough, Evan",
    title = "{Cosmological implications of Kalb-Ramond-like particles}",
    eprint = "2309.02485",
    archivePrefix = "arXiv",
    primaryClass = "hep-ph",
    doi = "10.1007/JHEP06(2024)075",
    journal = "JHEP",
    volume = "06",
    pages = "075",
    year = "2024"
}

@article{Jenks:2023pmk,
    author = "Jenks, Leah and Choi, Lyla and Lagos, Macarena and Yunes, Nicol{\'a}s",
    title = "{Parametrized parity violation in gravitational wave propagation}",
    eprint = "2305.10478",
    archivePrefix = "arXiv",
    primaryClass = "gr-qc",
    doi = "10.1103/PhysRevD.108.044023",
    journal = "Phys. Rev. D",
    volume = "108",
    number = "4",
    pages = "044023",
    year = "2023"
}

@article{Heisenberg:2014rta,
    author = "Heisenberg, Lavinia",
    title = "{Generalization of the Proca Action}",
    eprint = "1402.7026",
    archivePrefix = "arXiv",
    primaryClass = "hep-th",
    doi = "10.1088/1475-7516/2014/05/015",
    journal = "JCAP",
    volume = "05",
    pages = "015",
    year = "2014"
}

@article{Heisenberg:2019akx,
    author = "Heisenberg, Lavinia and Trenkler, Georg",
    title = "{Generalization of the 2-form interactions}",
    eprint = "1908.09328",
    archivePrefix = "arXiv",
    primaryClass = "hep-th",
    doi = "10.1088/1475-7516/2020/05/019",
    journal = "JCAP",
    volume = "05",
    pages = "019",
    year = "2020"
}

@article{Germani:2009iq,
    author = "Germani, Cristiano and Kehagias, Alex",
    title = "{P-nflation: generating cosmic Inflation with p-forms}",
    eprint = "0902.3667",
    archivePrefix = "arXiv",
    primaryClass = "astro-ph.CO",
    doi = "10.1088/1475-7516/2009/03/028",
    journal = "JCAP",
    volume = "03",
    pages = "028",
    year = "2009"
}

@article{Ajith:2022wia,
    author = "Ajith, Abhijith and Panda, Sukanta",
    title = "{Inflation using a triplet of antisymmetric tensor fields}",
    eprint = "2212.13508",
    archivePrefix = "arXiv",
    primaryClass = "gr-qc",
    doi = "10.1140/epjc/s10052-023-11932-x",
    journal = "Eur. Phys. J. C",
    volume = "83",
    number = "8",
    pages = "770",
    year = "2023"
}

@article{Emami:2016ldl,
    author = "Emami, Razieh and Mukohyama, Shinji and Namba, Ryo and Zhang, Ying-li",
    title = "{Stable solutions of inflation driven by vector fields}",
    eprint = "1612.09581",
    archivePrefix = "arXiv",
    primaryClass = "hep-th",
    doi = "10.1088/1475-7516/2017/03/058",
    journal = "JCAP",
    volume = "03",
    pages = "058",
    year = "2017"
}

@article{Esposito-Farese:2009wbc,
    author = "Esposito-Farese, Gilles and Pitrou, Cyril and Uzan, Jean-Philippe",
    title = "{Vector theories in cosmology}",
    eprint = "0912.0481",
    archivePrefix = "arXiv",
    primaryClass = "gr-qc",
    doi = "10.1103/PhysRevD.81.063519",
    journal = "Phys. Rev. D",
    volume = "81",
    pages = "063519",
    year = "2010"
}

@article{Potting:2023xzt,
    author = "Potting, Robertus",
    title = "{A stable Lorentz-breaking model for an antisymmetric two-tensor}",
    eprint = "2311.03159",
    archivePrefix = "arXiv",
    primaryClass = "hep-th",
    month = "11",
    year = "2023"
}

@article{Aashish:2021gdf,
    author = "Aashish, Sandeep and Ajith, Abhijith and Panda, Sukanta and Thakur, Rahul",
    title = "{Inflation with antisymmetric tensor field: new candidates}",
    eprint = "2112.11432",
    archivePrefix = "arXiv",
    primaryClass = "gr-qc",
    doi = "10.1088/1475-7516/2022/04/043",
    journal = "JCAP",
    volume = "04",
    number = "04",
    pages = "043",
    year = "2022"
}

@article{Altschul:2009ae,
    author = "Altschul, Brett and Bailey, Quentin G. and Kostelecky, V. Alan",
    title = "{Lorentz violation with an antisymmetric tensor}",
    eprint = "0912.4852",
    archivePrefix = "arXiv",
    primaryClass = "gr-qc",
    reportNumber = "IUHET-537",
    doi = "10.1103/PhysRevD.81.065028",
    journal = "Phys. Rev. D",
    volume = "81",
    pages = "065028",
    year = "2010"
}

@article{Adshead:2012kp,
    author = "Adshead, Peter and Wyman, Mark",
    title = "{Chromo-Natural Inflation: Natural inflation on a steep potential with classical non-Abelian gauge fields}",
    eprint = "1202.2366",
    archivePrefix = "arXiv",
    primaryClass = "hep-th",
    doi = "10.1103/PhysRevLett.108.261302",
    journal = "Phys. Rev. Lett.",
    volume = "108",
    pages = "261302",
    year = "2012"
}

@article{Seifert:2019kuz,
    author = "Seifert, Michael D.",
    title = "{Singular Hamiltonians in models with spontaneous Lorentz symmetry breaking}",
    eprint = "1903.06140",
    archivePrefix = "arXiv",
    primaryClass = "hep-th",
    doi = "10.1103/PhysRevD.100.065017",
    journal = "Phys. Rev. D",
    volume = "100",
    number = "6",
    pages = "065017",
    year = "2019"
}

@article{Adshead:2013nka,
    author = "Adshead, Peter and Martinec, Emil and Wyman, Mark",
    title = "{Perturbations in Chromo-Natural Inflation}",
    eprint = "1305.2930",
    archivePrefix = "arXiv",
    primaryClass = "hep-th",
    doi = "10.1007/JHEP09(2013)087",
    journal = "JHEP",
    volume = "09",
    pages = "087",
    year = "2013"
}

@article{Aashish:2019zsy,
    author = "Aashish, Sandeep and Padhy, Abhilash and Panda, Sukanta",
    title = "{Avoiding instabilities in antisymmetric tensor field driven inflation}",
    eprint = "1901.10959",
    archivePrefix = "arXiv",
    primaryClass = "gr-qc",
    doi = "10.1140/epjc/s10052-019-7308-0",
    journal = "Eur. Phys. J. C",
    volume = "79",
    number = "9",
    pages = "784",
    year = "2019"
}

@article{Aashish:2018lhv,
    author = "Aashish, Sandeep and Padhy, Abhilash and Panda, Sukanta and Rana, Arun",
    title = "{Inflation with an antisymmetric tensor field}",
    eprint = "1808.04315",
    archivePrefix = "arXiv",
    primaryClass = "gr-qc",
    doi = "10.1140/epjc/s10052-018-6366-z",
    journal = "Eur. Phys. J. C",
    volume = "78",
    number = "11",
    pages = "887",
    year = "2018"
}

@article{PhysRevD.9.2273,
  title = {Classical direct interstring action},
  author = {Kalb, Michael and Ramond, P.},
  journal = {Phys. Rev. D},
  volume = {9},
  issue = {8},
  pages = {2273--2284},
  numpages = {0},
  year = {1974},
  month = {Apr},
  publisher = {American Physical Society},
  doi = {10.1103/PhysRevD.9.2273},
  url = {https://link.aps.org/doi/10.1103/PhysRevD.9.2273}
}

@article{Terschlusen:2013iqa,
    author = {Terschl{\"u}sen, Carla and Strandberg, Bruno and Leupold, Stefan and Eichst{\"a}dt, Fabian},
    title = "{Reactions with pions and vector mesons in the sector of odd intrinsic parity}",
    eprint = "1305.1181",
    archivePrefix = "arXiv",
    primaryClass = "hep-ph",
    doi = "10.1140/epja/i2013-13116-6",
    journal = "Eur. Phys. J. A",
    volume = "49",
    pages = "116",
    year = "2013"
}

@article{Copeland:1984qk,
    author = "Copeland, Edmund J. and Toms, D. J.",
    title = "{Quantized Antisymmetric Tensor Fields and Selfconsistent Dimensional Reduction in Higher Dimensional Space-times}",
    reportNumber = "NCL 84-TP3",
    doi = "10.1016/0550-3213(85)90134-8",
    journal = "Nucl. Phys. B",
    volume = "255",
    pages = "201--230",
    year = "1985"
}

@article{Beekman:2010zx,
    author = "Beekman, A. J. and Sadri, D. and Zaanen, J.",
    title = "{Condensing Nielsen-Olesen strings and the vortex-boson duality in 3+1 and higher dimensions}",
    eprint = "1006.2267",
    archivePrefix = "arXiv",
    primaryClass = "cond-mat.str-el",
    doi = "10.1088/1367-2630/13/3/033004",
    journal = "New J. Phys.",
    volume = "13",
    pages = "033004",
    year = "2011"
}

@article{Ajith:2025rty,
    author = "Ajith, Abhijith and Panda, Sukanta",
    title = "{Scalar and vector modes in inflation with antisymmetric tensor field}",
    eprint = "2508.10159",
    archivePrefix = "arXiv",
    primaryClass = "gr-qc",
    doi = "10.1088/1475-7516/2025/10/086",
    journal = "JCAP",
    volume = "10",
    pages = "086",
    year = "2025"
}

@article{LIGOScientific:2018mvr,
    author = "Abbott, B. P. and others",
    collaboration = "LIGO Scientific, Virgo",
    title = "{GWTC-1: A Gravitational-Wave Transient Catalog of Compact Binary Mergers Observed by LIGO and Virgo during the First and Second Observing Runs}",
    eprint = "1811.12907",
    archivePrefix = "arXiv",
    primaryClass = "astro-ph.HE",
    reportNumber = "LIGO-P1800307",
    doi = "10.1103/PhysRevX.9.031040",
    journal = "Phys. Rev. X",
    volume = "9",
    number = "3",
    pages = "031040",
    year = "2019"
}

@article{LIGOScientific:2020ibl,
    author = "Abbott, R. and others",
    collaboration = "LIGO Scientific, Virgo",
    title = "{GWTC-2: Compact Binary Coalescences Observed by LIGO and Virgo During the First Half of the Third Observing Run}",
    eprint = "2010.14527",
    archivePrefix = "arXiv",
    primaryClass = "gr-qc",
    reportNumber = "P2000061",
    doi = "10.1103/PhysRevX.11.021053",
    journal = "Phys. Rev. X",
    volume = "11",
    pages = "021053",
    year = "2021"
}

@article{KAGRA:2021vkt,
    author = "Abbott, R. and others",
    collaboration = "KAGRA, VIRGO, LIGO Scientific",
    title = "{GWTC-3: Compact Binary Coalescences Observed by LIGO and Virgo during the Second Part of the Third Observing Run}",
    eprint = "2111.03606",
    archivePrefix = "arXiv",
    primaryClass = "gr-qc",
    reportNumber = "LIGO-P2000318",
    doi = "10.1103/PhysRevX.13.041039",
    journal = "Phys. Rev. X",
    volume = "13",
    number = "4",
    pages = "041039",
    year = "2023"
}

@article{LIGOScientific:2025slb,
    author = "Abac, A. G. and others",
    collaboration = "LIGO Scientific, VIRGO, KAGRA",
    title = "{GWTC-4.0: Updating the Gravitational-Wave Transient Catalog with Observations from the First Part of the Fourth LIGO-Virgo-KAGRA Observing Run}",
    eprint = "2508.18082",
    archivePrefix = "arXiv",
    primaryClass = "gr-qc",
    reportNumber = "LIGO-P2400386",
    month = "8",
    year = "2025"
}

@article{Jackiw:2003pm,
    author = "Jackiw, R. and Pi, S. Y.",
    title = "{Chern-Simons modification of general relativity}",
    eprint = "gr-qc/0308071",
    archivePrefix = "arXiv",
    reportNumber = "MIT-CTP-3409, BUHEP-03-18",
    doi = "10.1103/PhysRevD.68.104012",
    journal = "Phys. Rev. D",
    volume = "68",
    pages = "104012",
    year = "2003"
}

@article{Alexander:2009tp,
    author = "Alexander, Stephon and Yunes, Nicolas",
    title = "{Chern-Simons Modified General Relativity}",
    eprint = "0907.2562",
    archivePrefix = "arXiv",
    primaryClass = "hep-th",
    doi = "10.1016/j.physrep.2009.07.002",
    journal = "Phys. Rept.",
    volume = "480",
    pages = "1--55",
    year = "2009"
}

@article{Crisostomi:2017ugk,
    author = "Crisostomi, Marco and Noui, Karim and Charmousis, Christos and Langlois, David",
    title = "{Beyond Lovelock gravity: Higher derivative metric theories}",
    eprint = "1710.04531",
    archivePrefix = "arXiv",
    primaryClass = "hep-th",
    reportNumber = "LPT-ORSAY-18-33",
    doi = "10.1103/PhysRevD.97.044034",
    journal = "Phys. Rev. D",
    volume = "97",
    number = "4",
    pages = "044034",
    year = "2018"
}

@article{Nishizawa:2018srh,
    author = "Nishizawa, Atsushi and Kobayashi, Tsutomu",
    title = "{Parity-violating gravity and GW170817}",
    eprint = "1809.00815",
    archivePrefix = "arXiv",
    primaryClass = "gr-qc",
    reportNumber = "RUP-18-29",
    doi = "10.1103/PhysRevD.98.124018",
    journal = "Phys. Rev. D",
    volume = "98",
    number = "12",
    pages = "124018",
    year = "2018"
}

@article{Maleknejad:2011jw,
    author = "Maleknejad, A. and Sheikh-Jabbari, M. M.",
    title = "{Gauge-flation: Inflation From Non-Abelian Gauge Fields}",
    eprint = "1102.1513",
    archivePrefix = "arXiv",
    primaryClass = "hep-ph",
    doi = "10.1016/j.physletb.2013.05.001",
    journal = "Phys. Lett. B",
    volume = "723",
    pages = "224--228",
    year = "2013"
}

@article{Aoki:2025uwz,
    author = "Aoki, Katsuki and Fujita, Tomohiro and Kawaguchi, Ryodai and Yanagihara, Kazuki",
    title = "{Effective Field Theory of Chiral Gravitational Waves}",
    eprint = "2504.19059",
    archivePrefix = "arXiv",
    primaryClass = "astro-ph.CO",
    reportNumber = "WUCG-25-05, YITP-25-63",
    month = "4",
    year = "2025"
}

@article{Garriga:2025uko,
    author = "Garriga, Jaume and Gorji, Mohammad Ali and Hajkarim, Fazlollah and Sasaki, Misao",
    title = "{Oscillations and parity violation in gravitational wave background from extra tensor modes}",
    eprint = "2508.08481",
    archivePrefix = "arXiv",
    primaryClass = "astro-ph.CO",
    reportNumber = "YITP-25-90",
    month = "8",
    year = "2025"
}

@article{Alexander:2025wnj,
    author = "Alexander, Stephon and Daniel, Tatsuya and Manton, Tucker",
    title = "{Beyond general relativity: gravitational waves in non-minimally coupled theories}",
    eprint = "2510.25895",
    archivePrefix = "arXiv",
    primaryClass = "gr-qc",
    month = "10",
    year = "2025"
}

@article{Conroy:2019ibo,
    author = "Conroy, Aindri{\'u} and Koivisto, Tomi",
    title = "{Parity-Violating Gravity and GW170817 in Non-Riemannian Cosmology}",
    eprint = "1908.04313",
    archivePrefix = "arXiv",
    primaryClass = "gr-qc",
    doi = "10.1088/1475-7516/2019/12/016",
    journal = "JCAP",
    volume = "12",
    pages = "016",
    year = "2019"
}

@article{Bento:1992wy,
    author = "Bento, M. C. and Bertolami, O. and Moniz, P. V. and Mourao, J. M. and Sa, P. M.",
    title = "{On the cosmology of massive vector fields with SO(3) global symmetry}",
    eprint = "gr-qc/9302034",
    archivePrefix = "arXiv",
    reportNumber = "DFFCUL-03-5-1992, DF-IST-3-92, IFM-3-92",
    doi = "10.1088/0264-9381/10/2/010",
    journal = "Class. Quant. Grav.",
    volume = "10",
    pages = "285--298",
    year = "1993"
}

@article{Armendariz-Picon:2004say,
    author = "Armendariz-Picon, Christian",
    title = "{Could dark energy be vector-like?}",
    eprint = "astro-ph/0405267",
    archivePrefix = "arXiv",
    doi = "10.1088/1475-7516/2004/07/007",
    journal = "JCAP",
    volume = "07",
    pages = "007",
    year = "2004"
}

@article{Li:2020xjt,
    author = "Li, Mingzhe and Rao, Haomin and Zhao, Dehao",
    title = "{A simple parity violating gravity model without ghost instability}",
    eprint = "2007.08038",
    archivePrefix = "arXiv",
    primaryClass = "gr-qc",
    reportNumber = "USTC-ICTS/PCFT-20-19",
    doi = "10.1088/1475-7516/2020/11/023",
    journal = "JCAP",
    volume = "11",
    pages = "023",
    year = "2020"
}

@article{Li:2021wij,
    author = "Li, Mingzhe and Rao, Haomin and Tong, Yeheng",
    title = "{Revisiting a parity violating gravity model without ghost instability: Local Lorentz covariance}",
    eprint = "2104.05917",
    archivePrefix = "arXiv",
    primaryClass = "gr-qc",
    reportNumber = "USTC-ICTS/PCFT-21-17",
    doi = "10.1103/PhysRevD.104.084077",
    journal = "Phys. Rev. D",
    volume = "104",
    number = "8",
    pages = "084077",
    year = "2021"
}

@article{Li:2022mti,
    author = "Li, Mingzhe and Li, Zhihao and Rao, Haomin",
    title = "{Ghost instability in the teleparallel gravity model with parity violations}",
    eprint = "2201.02357",
    archivePrefix = "arXiv",
    primaryClass = "gr-qc",
    reportNumber = "USTC-ICTS/PCFT-22-01",
    doi = "10.1016/j.physletb.2022.137395",
    journal = "Phys. Lett. B",
    volume = "834",
    pages = "137395",
    year = "2022"
}

@article{Wu:2021ndf,
    author = "Wu, Qiang and Zhu, Tao and Niu, Rui and Zhao, Wen and Wang, Anzhong",
    title = "{Constraints on the Nieh-Yan modified teleparallel gravity with gravitational waves}",
    eprint = "2110.13870",
    archivePrefix = "arXiv",
    primaryClass = "gr-qc",
    doi = "10.1103/PhysRevD.105.024035",
    journal = "Phys. Rev. D",
    volume = "105",
    number = "2",
    pages = "024035",
    year = "2022"
}

@article{Hohmann:2022wrk,
    author = "Hohmann, Manuel and Pfeifer, Christian",
    title = "{Gravitational wave birefringence in spatially curved teleparallel cosmology}",
    eprint = "2203.01856",
    archivePrefix = "arXiv",
    primaryClass = "gr-qc",
    doi = "10.1016/j.physletb.2022.137437",
    journal = "Phys. Lett. B",
    volume = "834",
    pages = "137437",
    year = "2022"
}
\bibliographystyle{JHEP}
%-----------------------------------------------------------------%
\end{document}